\documentclass[10pt,final, conference]{IEEEtran}

\usepackage{amssymb}
\usepackage{pstricks}
\usepackage{amsmath}  
\usepackage[dvips]{graphicx}
\usepackage{colortbl}
\usepackage{enumerate}
\usepackage{ifthen}
\newtheorem{theorem}{Theorem}

\newtheorem{lemma}{Lemma}

\IEEEoverridecommandlockouts

\title{Performance of Multi-Antenna MMSE Receivers in Non-homogeneous Poisson Networks}

\author{\IEEEauthorblockN{Junjie~Zhu}
\IEEEauthorblockA{F. W. Olin College of Engineering\\
Needham, MA, USA\\
Email: junjie.zhu@students.olin.edu}
\and
\IEEEauthorblockN{Siddhartan Govindasamy$^\dagger$}
\IEEEauthorblockA{F. W. Olin College of Engineering\\
Needham, MA,  USA\\
Email: siddhartan.govindasamy@olin.edu}
\thanks{\noindent $\dagger$ Corresponding author. This research was supported in part by the National Science Foundation under Grant CCF-1117218.}}

\begin{document}

\maketitle 

\pagenumbering{arabic}

\begin{abstract}
A technique to compute the Cumulative Distribution Function (CDF) of the Signal-to-Interference-plus-Noise-Ratio (SINR) for a wireless link with a multi-antenna, Linear, Minimum-Mean-Square-Error (MMSE) receiver in the presence of interferers distributed according to a non-homogenous Poisson point process on the plane, and independent Rayleigh fading between antennas is presented. This technique is used to compute the CDF of the SINR for several different models of intensity functions, in particular, power-law intensity functions, circular-symmetric Gaussian intensity functions and intensity functions described by a polynomial in a bounded domain. Additionally it is shown that if the number of receiver antennas is scaled linearly with the intensity function, the SINR converges in probability to a limit determined by the ``shape'' of the underlying intensity function.  This work generalizes known results for homogenous Poisson networks to non-homogenous Poisson networks. 
\end{abstract}


\section{Introduction} 

Antenna arrays can improve the performance of wireless networks by increasing robustness through diversity, and data rates through spatial multiplexing, beamforming and  interference mitigation. The performance of multi-antenna systems in networks depends on environmental conditions and inter-node distances which effect signal and interference strengths and thus data rates. Hence the performance of multi-antenna systems in spatially distributed networks have received significant attention in the literature. 

Multi-antenna receivers in networks with uniformly random spatial node distribution have been studied in several works.  Govindasamy et. al. \cite{govindasamy2007spectral}  used an asymptotic analysis to approximate the spectral efficiency with MMSE receivers, and Jindal et. al. \cite{jindal2011multi} considered a partial zero-forcing receiver and found that it is possible to linearly increase the area spectral efficiency by simultaneously increasing the number of antennas and density of simultaneous transmissions. Ali et. al. \cite{ali2010performance} and Louie et. al. \cite{louie2011open} found the exact CDF of the SINR with MMSE receivers, with the former considering single-stream and the latter considering multi-stream transmissions. 

In many systems however, spatial node distributions may not be homogenous, such as in networks with hot-spots. Interference modeling in non-homogenous single antenna systems have been studied in several works such as \cite{WinNetworkInterference}, \cite{GantiClustered}, \cite{GantiHighSIR} and references therein. Multi-antenna systems in non-homogenous networks have been studied in relatively fewer works such as \cite{TreschCluster} who considered interference-alignment in clustered wireless networks, \cite{HunterMIMOCSMA} who considered multi-antenna systems in networks with Carrier-Sensing-Multiple-Access (CSMA) which induces correlation between actively transmitting nodes and \cite{NonHomogAsymp} who used an asymptotic analysis to analyze the spectral efficiency of non-homogenous networks with linear MMSE receivers. 
\begin{figure}[t]
\center
\includegraphics[width = 2.5in]{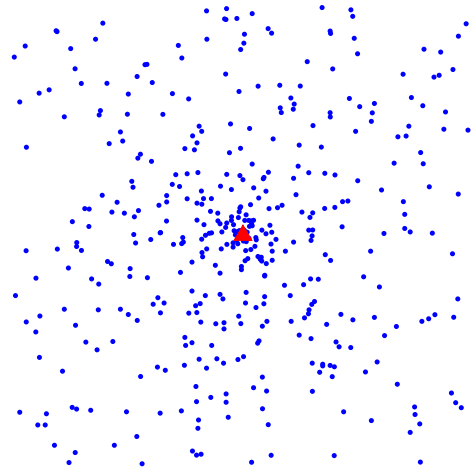}
\caption{Randomly distributed transmitters with intensity function $\Lambda(r,\theta) = 0.1r^{-1}$, with the representative receiver at the origin of the network.}
\label{Fig:NodesInPlane}
\end{figure}

Here, we develop a framework to characterize the SINR of a representative link with multiple antennas at the receiver in the presence of single-antenna interferers distributed according to a non-homogenous Poisson Point Process on the plane, and independent Rayleigh fading between all antennas. The non-homogenous node distribution is modeled  by an intensity function $\Lambda(r, \theta)$ which controls the likelihood of nodes occuring in a small region around the point $(r, \theta)$. Since $\Lambda(r, \theta)$ is deterministic (which can arise if the non-homogenity is predictable), this model differs works such as \cite{GantiClustered} and \cite{TreschCluster} where cluster locations are random. 

This framework is used to find a closed-form expression for the CDF of the SINR of a link in the center of a cluster with a power-law distribution of node intensities (analyzed asymptotically in \cite{NonHomogAsymp}) as shown in the example in Figure \ref{Fig:NodesInPlane} where $\Lambda(r, \theta) = 0.1r^{-1}$. Expressions involving generalized functions are also provided for networks with circularly-symmetric Gaussian intensities and  intensity functions  represented by a polynomial within a bounded region. The latter model is  interesting as arbitrary continuous intensity functions can be approximated within bounded intervals with arbitrary accuracy using polynomials. By fitting polynomials to experimental data, this result can also be used to characterize systems for which node distributions are not well-modeled mathematically but for which experimental data on node locations  exist.  Additionally, we show that if the number of receiver antennas $L$ is scaled linearly with $\Lambda(r, \theta)$, the SINR converges in probability to a positive deterministic value which is a function of $\Lambda(r, \theta)$. This latter result indicates that to the extent that our assumptions (in particular independent Rayleigh fading) hold, increasing the number of antennas per receiver can help scale such networks. The results here are derived combining techniques developed for homogenous networks in \cite{ali2010performance} and non-homogenous network models from \cite{NonHomogAsymp}.

\section{System Model}
In a circular network of radius $R$, a receiver centered at the origin is surrounded by $n$ transmitting nodes that are distributed independently and randomly in the circle according to the  Probability Density Function (PDF), $f_{r,\theta}(r,\theta)$, which is related to the intensity function $\Lambda(r,\theta)$ as follows:
\begin{align}\label{PDFtoIndFunc}
f_{r,\theta}(r,\theta) = \frac{r}{\mu}\Lambda(r,\theta)\mathbf{1}_{\{0 \le r < R\}}
\end{align}
where $\mathbf{1}_{\{0 \le r < R\}}$ is the indicator function. The number of the transmitters $n$ is a Poisson random variable with mean $\mu$, which can be expressed as: 
\begin{align}   \label{Eqn:Mean}
\mu = \int_0^R\int_0^{2\pi}r\Lambda(r,\theta)d\theta dr
\end{align}
We assume that these transmitters, also referred to as interferers, are communicating with other receivers at locations that do not affect our results. In addition to the interferers, a target transmitter is located at a fixed distance $r_T$ away from the receiver at the origin. 

Loss in signal power due to propagation is modeled by the inverse power-law model, so that the average power $p$ from a node transmitting with unit power, received at a distance $r$ is $p = r^{-\alpha}$, with the path-loss exponent $\alpha>2$. The receiver at the origin has $L$ antennas and each transmitter has one antenna. We use the subscript $T$ to denote the target transmitter whereas the interferers are labeled $1, 2, \cdots, n$. Using this notation, the transmitted symbol of the target transmitter is $x_T$ and the transmitted symbol of the $i$-th interferer is $x_i$. The $L\times 1$ vector $\mathbf{g}_T$ represents the channel coefficients between the target transmitter and the $L$ antennas of the receiver at the origin. Similarly, $\mathbf{g}_i$ and $r_i$ represent the channel coefficients and the distance between the $i$-th interferer and the receiver at the origin. The $L\times 1$ received signal vector can be described by the following equation:
\begin{align}
\mathbf{y} = r_T^{-\alpha/2}\mathbf{g}_T x_T + \sum_{i=1}^n r_i^{-\alpha/2}\mathbf{g}_i x_i + \mathbf{w}
\end{align}
where the entries of $\mathbf{g}_T$ and $\mathbf{g}_i$ are independent, identically-distributed (i.i.d.), zero-mean, unit-variance complex Gaussian random variables. $\mathbf{w}$ is a noise vector of zero-mean, i.i.d. complex Gaussian entries with variance $\sigma^2$ per complex dimension. 

The receiver uses a minimum mean square error (MMSE) estimator to estimate $x_T$ from $\mathbf{y}$. The MMSE estimator is known to maximize the SINR which is given by the following well-known formula
\begin{align}
\text{SINR} = r_T^{-\alpha}\mathbf{g}_T^\dagger \left( \mathbf{G}\mathbf{P}\mathbf{G}^{\dagger}+\sigma^2 \mathbf{I}_L \right)^{-1}\mathbf{g}_T
\end{align}
where the $i$-th column of the $L\times n$  matrix $\mathbf{G}$ represents the channel vector $\mathbf{g}_i$ of the $i$-th interferer. $\mathbf{I}_L$ is an $L \times L$ identity matrix, and $\mathbf{P} = diag\left[r_1^{-\alpha}, r_2^{-\alpha}, \dotsb r_n^{-\alpha}\right]$ is a diagonal matrix with entries corresponding to the received power from the interferers. Furthermore, we define the distance-normalized SINR as  $\gamma = \text{SINR}\cdot r_T^\alpha$ to simplify notation.
In deriving the main results we take $R \to \infty$ to model the interferers as resulting from a non-homogeneous Poisson point process with the intensity function $\Lambda(r, \theta)$.

\section{Main Results}

\subsection{Outage Probability with General Intensity Function}  

In communication systems, it is desirable to know the probability that the SINR is below a threshold $\tau$, which is often referred to as the outage probability. For a given $r_T$, this probability is simply $\text{Pr}\{\text{SINR} \le \tau\}= F_\gamma(\tau r_T^\alpha)$, where $F_\gamma(\gamma)$ is the CDF of $\gamma$ given by the following theorem.
\begin{theorem}\label{Theorem:GenCDF}
\begin{eqnarray} \label{Eqn:GenCDF}
 F_\gamma(\gamma) &=&  1-\sum_{k=0}^{L-1}\frac{(\psi(\gamma)+\sigma^2\gamma)^k}{k!}\exp(-\psi(\gamma)-\sigma^2\gamma) \nonumber \\
&=& 1-\frac{\Gamma(L, \psi(\gamma)+\sigma^2 \gamma)}{\Gamma(L)}
\end{eqnarray}
where
\begin{align} 
\psi(\gamma) =  \int_0^\infty \int_0^{2\pi} \Lambda(r,\theta)r\frac{r^{-\alpha}\gamma}{1+r^{-\alpha}\gamma}d\theta dr, \label{Eqn:PsiDef}
\end{align}
and, $\Gamma(.)$ and $\Gamma(.,.)$ are the gamma function and the upper incomplete gamma function respectively.
\end{theorem}
 The corresponding PDF is:
\begin{align*}
 f_\gamma(\gamma) =  \frac{(\psi(\gamma)+\sigma^2\gamma)^{L-1}\exp(-\psi(\gamma)-\sigma^2\gamma)(\sigma^2+\psi^{\prime} (\gamma))}{\Gamma(L)},
\end{align*}
where $\psi^{\prime} (\gamma)$ is the first derivative of  $\psi(\gamma)$ with respect to $\gamma$.
{\it Proof:} Given in Appendix \ref{Sec:ProofOfGenCDF}.

As a result, increasing the number of antennas at the receiver from $L$ to $L+1$ reduces the outage probability by $(\psi(\gamma)+\sigma^2\gamma)^{L}\exp(-\psi(\gamma)-\sigma^2\gamma)/ L!$.

\subsection{Scaling Non-homogeneous Networks by Increasing the Number of Antennas} \label{Sec:Scaling}
In this section, we show that the SINR on the representative link (link between the receiver at the origin and its target transmitter) converges to a constant as the density of nodes increases, if the number of antennas at the receiver is linearly increased with interferer density. This is under the assumption that the channel model holds (in particular the independent Rayleigh fading assumption) and that accurate measurements of channel state information are available at the receiver.
A similar problem was addressed in the context of homogeneous networks in \cite{govindasamy2007spectral} and  \cite{jindal2011multi}. 

A key result that we use to prove this is the following lemma which may already be known but we were not able to find it in the literature.
\begin{lemma} \label{Lemma:RegGammaLimit}
Let the upper regularized gamma function  be denoted by  $Q(L,x)=\frac{\Gamma(L,x)}{\Gamma(L)}$, where $\Gamma(L, x)$ is the upper incomplete gamma function.  Let $L$ be a positve integer and $q > 0$, then
\begin{align} \label{Eqn:Lemma1}
\lim_{L\to\infty} Q(L, q L) = \begin{cases} 0, & \text{if $q \ge1$}
\\
1, &\text{if $q < 1$}.
\end{cases}
\end{align}
\end{lemma}
{\it Proof:} Given in Appendix \ref{Sec:ProofOfRegGammaLimit}.

Suppose that the intensity function $\Lambda(r,\theta) = \beta\Lambda_c(r,\theta)$, where $\Lambda_c(r,\theta)$ is a nominal intensity function which describes the "shape" of the true intensity function, and   $\beta$ is the nominal interferer density which scales the nominal intensity function. We also define: 
\begin{align} \label{Eqn:PsiC}
\psi_c(\gamma) =   \int_0^\infty \int_0^{2\pi} \Lambda_c(r,\theta)r\frac{r^{-\alpha}\gamma}{1+r^{-\alpha}\gamma}d\theta dr.
\end{align}
Suppose that $\beta = qL$ , i.e., the number of antennas is scaled linearly with the nominal density,  we have $\psi(\gamma) = \beta\psi_c(\gamma) = qL\psi_c(\gamma)$. In addition, assuming that the noise is negligible, the CDF in \eqref{Eqn:GenCDF} from Theorem \ref{Theorem:GenCDF} can be expressed as:
\begin{align}
F_\gamma(\gamma) = 1-\frac{\Gamma(L, qL\psi_c(\gamma))}{\Gamma(L)} 
\end{align} 
By Lemma \ref{Lemma:RegGammaLimit}, we have:
\begin{align*}
 \lim_{L \to \infty}F_\gamma(\gamma) = 1-\frac{\Gamma(L, qL\psi_c(\gamma))}{\Gamma(L)} = \begin{cases} 0,  & \text{if $\gamma <\psi_c^{-1}\left(\frac{1}{q}\right) $}      \\     1, &\text{if $\gamma \ge \psi_c^{-1}\left(\frac{1}{q}\right)$}.
\end{cases}
\end{align*}
which implies that the SINR converges in distribution to a constant $\psi_c^{-1}\left(\frac{1}{q}\right)r_T^{-\alpha}$ as the number of antennas goes to infinity. Additionally, since convergence in distribution to a constant implies convergence in probability (e.g. see \cite{Karr}), $\gamma$ converges in probability as well. Therefore, if we increase the number of antennas linearly with the nominal interferer density in a given network, the SINR will approach a constant non-zero value. This fact implies that such networks can be scaled by linearly increasing the number of antennas per receiver with user density without degrading the SINR to zero, provided that the assumptions are satisfied.

\subsection{Outage Probability with Polynomial Intensity Function} \label{Sec:PolyInt}
Consider an intensity function of the following form:
\begin{align} \label{Eqn:Poly}
\Lambda_m^P(r, \theta) = \sum_{k=0}^m a_k r^k  \textbf{1}_{\{0 \le r \le R_0\}}+\rho_0 r^\epsilon \textbf{1}_{\{R_0\le r\}},
\end{align}
where $a_0$, $a_1$, $a_2$, ...$a_m$ are arbitrary polynomial coefficients. Thus, the intensity function in the range $0<r<R_0$ is described by a polynomial. The second term in the equation results in a power-law decay of the interferer density for $r \ge R_0$. We additionally assume that $-2< \epsilon<-1$ and $\rho_0>0$.
Then, the corresponding CDF of $\gamma$ is:
\begin{equation}\label{FNP}
 F_m^P(\gamma)  =  1-\sum_{i=0}^{L-1}\frac{(\psi_m^P(\gamma)+\sigma^2\gamma)^i}{i!}\exp(-\psi_m^P(\gamma)-\sigma^2\gamma)
\end{equation}
where 
\begin{align}
&\psi_m^P(\gamma) =  \frac{2\pi^2\rho_0\gamma^{\frac{2+\epsilon}{\alpha}}}{\alpha}\csc \left( \frac{(2+\epsilon)\pi}{\alpha} \right) +  \nonumber \\
&\sum_{k=0}^m\frac{ 2\pi a_k R_0^{2+k}}{2+k}{_2F_1}\!\bigg( 1, \frac{2+k}{\alpha}; \frac{2+\alpha + k}{\alpha} -\frac{R_0^\alpha}{\gamma} \! \bigg),
\end{align}
where $_2F_1(\cdot,\cdot;\cdot;\cdot)$ is  Gauss's hypergeometric function. 

Since any continuous function can be uniformly approximated by a polynomial function with arbitrary accuracy in a bounded interval, according to the Stone-Weierstrass Theorem \cite{Rudin}, we use the result above to derive the following theorem. 

\begin{theorem} \label{Theorem:PolyIntensity}
For any CDF $F(\gamma)$ corresponding to the intensity function
\begin{align} \label{Eqn:ArbIndFunc}
\Lambda(r, \theta) =h(r)  \textbf{1}_{\{0<r<R_0\}}+\rho_0 r^\epsilon \textbf{1}_{\{R_0 \le r\}}
\end{align}
where $h(r)$ is any continuous function of $r$, there exist coefficients $a_0$, $a_1$, $a_2$, ...$a_m$ such that 
\begin{align}
\lim_{m \to \infty}F_m^P(\gamma) = F(\gamma).
\end{align}
\end{theorem}
{\it Proof:} Given in Appendix \ref{Sec:ProofOfPolyIntensity}.

Theorem 2 allows us to approximate with arbitrary accuracy, the CDF of the SINR for any intensity function that is continuous in $r$ within a finite domain using a polynomial expression. In particular, since efficient algorithms exist for fitting polynomials to real data, this technique could be useful to analyze networks whose geometrical characteristics are not easily captured by mathematical models, but for which experimental data on node positions are available.

\subsection{Outage Probability with Piecewise Power-law Function}
Consider an intensity function of the following form:
	\begin{align} \label{Eqn:PiecewiseInt}
	\Lambda(r,\theta) = \begin{cases}
 	\rho_1r^{\epsilon_1} & \text{for  $0<r<R_1$} \\
 	\rho_2r^{\epsilon_2} & \text{for $R_1<r<R_2$} \\
            ...\\
	\rho_m r^{\epsilon_m} & \text{for $R_{m-1}<r<R_m$}\\
	\end{cases}
	\end{align}
where $\epsilon_k > - 2$ for each $k$ for which $R_k = 0$. With this intensity function, in the range $R_{k-1} < r \leq R_{k}$, the intensity function of the interferers follows a power-law distribution with nominal density $\rho_k$, and exponent $\epsilon_k$. The CDF of $\gamma$ in this case is given by \eqref{Eqn:GenCDF} with 
\begin{align}
&\psi(\gamma)=   \frac{2\pi \rho_1 R_1^{2+\epsilon_1}}{2+\epsilon_1}{_2F_1}\bigg( 1, \frac{2+\epsilon_1}{\alpha}; \frac{2+\alpha + \epsilon_1}{\alpha} ;  -\frac{R_1^\alpha}{\gamma} \bigg) \nonumber \\
& + \sum_{k=2}^{m-1} \frac{2\pi \rho_k \gamma}{2-\alpha+\epsilon_k} \bigg [   R_k^{2-\alpha+\epsilon_k} \times \nonumber \\
& {_2F_1}\bigg(  1,  \frac{-2+\alpha-\epsilon_k}{\alpha}; \frac{-2+2\alpha-\epsilon_k}{\alpha}; -\gamma R_k^{-\alpha}  \bigg)  -   R_{k-1}^{2-\alpha+\epsilon_k}  \nonumber \\ 
& \times {_2F_1}\!\bigg(\!1, \frac{-2+\alpha-\epsilon_k}{\alpha};\frac{-2+2\alpha-\epsilon_k}{\alpha}; -\gamma R_{k-1}^{-\alpha}  \bigg) \bigg ]  \nonumber \\
&- \frac{2\pi \rho_m \gamma}{2-\alpha+\epsilon_m}  R_{m-1}^{2-\alpha+\epsilon_m} \times \nonumber \\
&{_2F_1}\!\bigg(\!  1,   \frac{-2+\alpha-\epsilon_m}{\alpha};\frac{-2+2\alpha-\epsilon_m}{\alpha}; -\gamma R_{m-1}^{-\alpha}  \bigg).
\end{align}
For the simplest case, consider the intensity function:
\begin{align}\label{Eqn:PowerLaw}
\Lambda(r,\theta) = \rho r^\epsilon,
\end{align}
where $ -2<\epsilon < \alpha - 2$, to prevent interference from going to infinity as $R \to \infty$. This intensity function can be used to model a network with a dense cluster of interferers, centered on the representative receiver and is useful to model networks with hot-spots. In this case, evaluating $\psi(\gamma)$ over a finite $R$, applying one of Euler's hypergeometric transforms \cite{AbramovitzStegun}, and taking $R\to\infty$ yields, 
\begin{eqnarray}  \label{Eqn:PLPsi}
	\psi(\gamma) = \frac{2\pi^2\rho}{\alpha} \gamma^{(\epsilon+2)/\alpha}\csc\left(\pi\frac{\epsilon+2}{\alpha}\right).
 	\end{eqnarray}
Substituting \eqref{Eqn:PLPsi} into \eqref{Eqn:GenCDF}, we have the CDF of $\gamma$ as
\begin{align} 
F_\gamma(\gamma) &=  1-\sum_{i=0}^{L-1} \frac{2\pi^2\rho}{i! \alpha} \csc\left(\pi\frac{\epsilon+2}		 			{\alpha}\right)\gamma^{(\epsilon+2)/\alpha}+\sigma^2\gamma)^i \times \quad \quad \nonumber \\
	& \exp\left(- \frac{2\pi^2\rho}{\alpha} 	\csc\left(\pi\frac{\epsilon+2}{\alpha}\right)\gamma^{(\epsilon+2)/\alpha}-\sigma^2\gamma\right) 
\end{align}

\subsection{Outage Probability with Gaussian Intensity Function}
Suppose that the receiver is located in the center of a cluster who intensity follows a circularly-symmetric Gaussian function.  $\Lambda(r, \theta)$ can be written in the form of the PDF of a Rayleigh variable  multiplied with a constant $\rho$ expressed as:
	\begin{align}
	\Lambda(r,\theta)= \rho\frac{r}{v^2}e^{-r^2/2v^2}
	\end{align}
where $v$  controls the width of the intensity function. For specific integer values of  $\alpha$,  $\psi(\gamma)$ can be evaluated in terms of generalized functions such as the hypergeomtric and Meijer-G functions. For instance, when $\alpha = 3$,
	\begin{align} 
	\psi(\gamma) =  \frac{\gamma \rho }{2 \sqrt{3} \pi  v^2}G^{5\  2}_{2\ 5 } \left( \frac{\gamma^2}{216 v^6} \bigg|  \begin{matrix}0, \frac{1}{2} \\ 0,0,\frac{1}{3},\frac{1}{2},\frac{2}{3}\end{matrix} \right) \label{GaussianIntensitya3}
	\end{align}
where $G$ is the Meijer G-function. These expressions involve special functions which can be evaluated efficiently in most mathematical software packages.

\section{Numerical Results}
\subsection{Monte-Carlo Simulations} 
In this section, we summarize numerical results for the cases analyzed in the previous sections. Monte-Carlo simulations were run for various intensity functions to validate the general technique in Theorem \ref{Theorem:GenCDF}. 

\begin{figure}[htbp]
\includegraphics[width =3.25in]{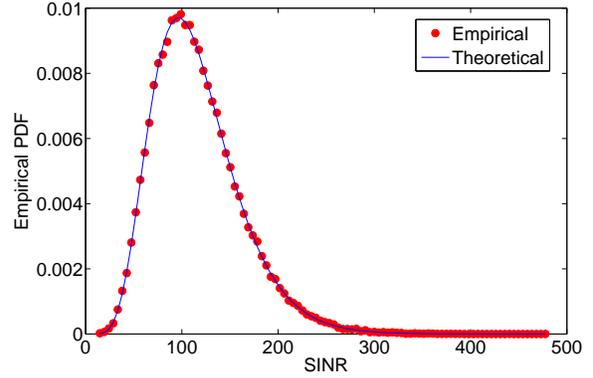}
\caption{Comparison between the empirical and theoretical PDF of the SINR for the Gausian intensity function with  mean number of interferers $\mu=1000$,  $v = 500$ ,  $r_T = 20$, number of receiver antennas $L = 10$, $\alpha = 3$, $\sigma^2 = 10^{-14}$. }
\label{Fig:PDFMatch}
\end{figure}

Figure \ref{Fig:PDFMatch} shows the empirical probability density function (PDF) from 100,000 simulations of a wireless network with a circularly-symmetric Gaussian intensity function, along with a graph of the PDF from Theorem 1. The additional parameters of the simulations are given in the caption. From the graph, it is clear that the simulations match the theoretical predictions. Additionally note that even with a large number of trials, the simulated PDF is not smooth which suggests that a purely simulation based approach to estimate the PDF of the SINR in this case is computationally prohibitive. Hence, the CDF and PDF  given in terms of generalized functions through \eqref{GaussianIntensitya3} are useful as they can be evaluated efficiently.

\begin{figure}
\includegraphics[width = 3.25in]{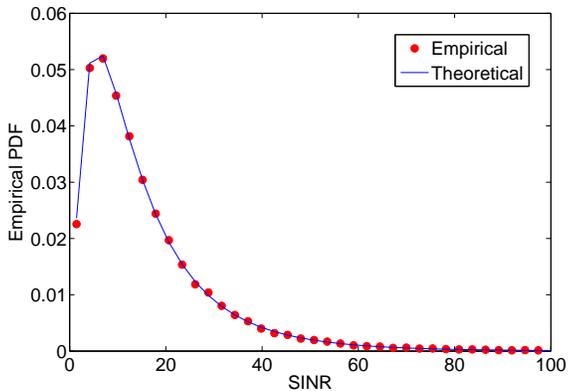} 
\caption{Comparison between the empirical and theoretical probability density function of SINR with the power-law intensity function $\Lambda(r,\theta) = \frac{0.023 }{\sqrt{r}}$ . The parameters used are $r_T = 10$, $L = 10$,  $\alpha = 4$, $\sigma^2 = 10^{-12}$, and $100,000$  Monte-Carlo trials.}
\label{Fig:PowerLaw}
\end{figure}

Figure \ref{Fig:PowerLaw} shows the simulated and theoretical PDF for the power-law intensity function $\Lambda(r,\theta) = \frac{0.023 }{\sqrt{r}}$. The additional parameters of the simulation are shown in the figure caption. From the graph, it is clear that the theoretical prediction of Theorem 1 is accurate.

\subsection{Uniform Versus Clustered Networks}
We can use the power law intensity function of \eqref{Eqn:PowerLaw} with different values of the exponent $\epsilon$ to compare the SINRs between networks with uniform and clustered node distributions.  

\begin{figure}[htbp]
\includegraphics[width = 3.25in]{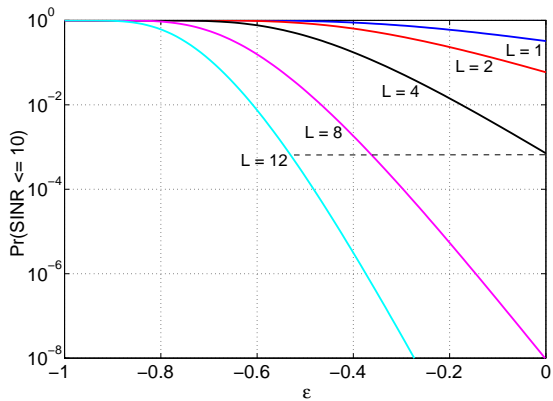}
\caption{Outage probability vs. $\epsilon$ for power-law intensity functions with $L \! = \! 1, 2,4,8$ and $12$ receiver antennas. Representative link length $r_T = 5$, mean number of interferers in a radius 1000 circle is fixed at $\mu=3142$,  $\alpha = 4$, and $\sigma^2 = 10^{-12}$.}
\label{Fig:ConstantR_E}
\end{figure}

Note that if we keep the value of $\rho$ fixed, different values of $\epsilon$ result in radically different numbers of nodes in the vicinity of the receiver at the origin.  To make a fairer comparison between different values of $\epsilon$, we adjust $\rho$ such that the mean  number of nodes that fall in a radius $R_c$ disk centered at the origin of the infinite network is fixed for the values of $\epsilon$ under consideration. Figure \ref{Fig:ConstantR_E} shows the probability that the SINR is less than or equal to 10 for different values of $\epsilon$. For this plot, we have assumed that $\mu = 3142$ nodes on average in a circle of radius $R_c = 1000$, with $\epsilon$ varying from $-1$ to $0$, and values of $\rho$ selected so that $\mu = 3142$. The remaining parameters are specified in the caption. 
Note that the outage probability increases significantly with clustering. For instance, with $L  =  4$, the outage probability is just below $10^{-3}$ for $\epsilon = 0$, and for $\epsilon = -0.5$, the outage probability is greater than 0.1. Additionally, observe that it is possible to significantly reduce the outage probability by increasing the number of antennas at the receiver since $L = 12$ antennas at the receiver with $\epsilon \approx -0.5$ has the same outage probability as $L=4$ in a homogenous network. 

\subsection{Scaling node density by increasing number of antennas}
\begin{figure}[htp]
\includegraphics[width =3.25in]{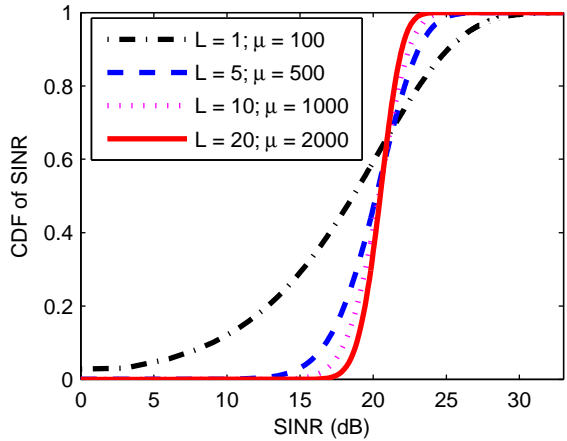} 
\caption{Cumulative distribution function of SINR (dB) with number of antennas increasing linearly with nominal interferer density, with $v = 500$ , $r_T = 20$,$\alpha = 3$ and  $\sigma^2 = 10^{-14}$.}
\label{Fig:scaled_rho}
\end{figure}

 Section \ref{Sec:Scaling} shows that if the number of antennas is scaled linearly with the intensity function, the SINR approaches a deterministic, non-zero value. To verify this result, we plotted the CDF of the SINR for the Gaussian intensity function with $\Lambda(r, \theta) = \beta \Lambda_c(r,\theta)  = \beta \frac{r}{v^2}e^{-r^2/2v^2}$. $L = 1, 5, 10$ and $20$ are considered with $\beta$ increasing linearly with $L$ in Figure \ref{Fig:scaled_rho}.  The remaining parameters used for the plot are given in the caption. Figure \ref{Fig:scaled_rho} shows that as the number of interferers increases from $1$ to $20$, the CDF of SINR approaches a step function, i.e the SINR approaches a constant non-zero value in distribution implying that it converges in probability as well. 

\section{Summary and Conclusions}
A technique to compute the CDF of the SINR on a link with multiple receiver antenans in non-homogenous Poisson field of interferers is presented and used to find expressions for the CDF of the SINR for several representative node distributions including  power-law intensity functions for which a closed form expression for the CDF is found. This result can be used to characterize the SINR in the center of a cluster and indicates that while the SINR is significantly smaller in the center of a dense cluster than in a homogenous network, the performance loss can be mitigated by using a larger number of antennas at the cost of additional complexity. Additionally, we showed that if the number of receiver antennas is scaled linearly with the node intensity function, the SINR converges in probability to a positive constant, indicating that it is possible to scale such networks by increasing the number of antennas provided that the system assumptions  hold. These results are useful to designers of wireless systems where the spatial distribution of nodes are predictable, a situation that often arises in practice. 

\appendix

\subsection{Proof of Theorem \ref{Theorem:GenCDF}}\label{Sec:ProofOfGenCDF}
From \cite{ali2010performance}, the CDF of $\gamma$  can be expressed as
\begin{align}\label{longeqn}
&F_\gamma(\gamma) =1-\exp(-\sigma^2\gamma)\sum_{i=0}^{L-1}\sum_{k=0}^i \frac{(\sigma^2\gamma)^{i-k}}{k!(i-k)!} \cdot \nonumber \\
&\;\;\;\;\;\left(\mu\text{E}_{p}\left[ \frac{ p\gamma}{1+p\gamma}\right]   \right)^k  \exp\left(-\mu\text{E}_{p} \left[ \frac{ p\gamma}{1+p\gamma}\right]   \right), 
\end{align}
where $\text{E}_x$ represents the expectation with respect to the random variable $x$. Recall that $p = r^{-\alpha}$ and the locations of the transmitting nodes are characterized by the PDF $f_{r,\theta}(r,\theta)$. Consequently, for our network model we have the following expressions for the expectation in the previous equation:
\begin{align}
\text{E}_{p}\left[ \frac{ p\gamma}{1+p\gamma}\right]  &= \lim_{R\to\infty}\int_0^R\int_0^{2\pi} f_{r,\theta}(r,\theta)\frac{r^{-\alpha}\gamma}{1+r^{-\alpha}\gamma}d\theta dr  \label{E1} 
\end{align}
From \eqref{PDFtoIndFunc} and \eqref{Eqn:PsiDef}, we have
\begin{align}
\psi(\gamma)=\mu \text{E}_{p}\left[ \frac{ p\gamma}{1+p\gamma}\right]
\end{align} 
Substituting  into \eqref{longeqn} yields
\begin{align}
 &F_\gamma(\gamma) = 1-\exp(-\sigma^2\gamma)\sum_{i=0}^{L-1}\sum_{k=0}^i\frac{(\sigma^2\gamma)^{i-k}}{k!(i-k)!}\psi^k(\gamma) \exp(\psi(\gamma))  \nonumber \\
&=  1-\sum_{i=0}^{L-1}\frac{(\psi(\gamma)+\sigma^2\gamma)^i}{i!}\exp(-\psi(\gamma)-\sigma^2\gamma) 
\end{align}
\eqref{Eqn:GenCDF} follows from equation (6.5.13) in \cite{AbramovitzStegun}, and the PDF is found by taking the derivative of the CDF and simplifying.

\subsection{Proof of Lemma \ref{Lemma:RegGammaLimit}} \label{Sec:ProofOfRegGammaLimit}
Let  $X_0, X_1,\cdots, X_{L-1}$ be a set of i.i.d. Poisson random variables with mean $q$. Define their sum and average respectively by $Y = \sum_{k = 0}^{L-1} X_k$ and $\bar{Y} = \frac{Y}{L}$. By the weak law of large numbers $\bar{Y} \to q$ as $L \to \infty$ in probability, which implies that  $\bar{Y} \to q$ in distribution, i.e. 
\begin{align}
\lim_{L\to\infty} \Pr\left(\bar{Y} \leq x \right) = \begin{cases} 0, & \text{if $x\leq q$}
\\
1, &\text{if $x > q$}.
\end{cases}\label{Eqn:NormalizedPoissonSum}
\end{align}
Since the sum of independent Poisson random variables is another Poisson random variable, $Y$ is a Poisson random variable with mean $qL$. Thus,
\begin{align}
Q(L, qL) = \Pr\left(Y \leq L\right) = \Pr\left(\bar{Y} \leq 1\right).
\end{align}
where $Q(L, qL)$ is  the CDF of a Poisson random variable with mean $qL$.  Taking the limit as $L \to\infty$ and substituting \eqref{Eqn:NormalizedPoissonSum} yields \eqref{Eqn:Lemma1} completing the proof.

\subsection{Proof of Theorem \ref{Theorem:PolyIntensity}}\label{Sec:ProofOfPolyIntensity}
According to the Stone-Weierstrass Theorem \cite{Rudin}, for every $\delta>0$ there exists $\Lambda_m^P(r, \theta)$ such that for all $r$ in $[0,R]$, $\exists$ an integer $N$ such that $m \ge N$ implies:
\begin{align}
\left|  \Lambda_m^P(r, \theta)- \Lambda(r, \theta)  \right|  < \delta
\end{align}
Let $\delta_1$ be the product of $\delta$ and the maximum value of $r\frac{r^{-\alpha}\gamma}{1+r^{-\alpha}\gamma}$,
\begin{align}
\max\left[r\frac{r^{-\alpha}\gamma}{1+r^{-\alpha}\gamma}\right] = \left( \frac{\gamma}{\alpha-1}\right)^{1/\alpha} \frac{(\alpha-1)}{\alpha} > 0 
\end{align}
Then, for any $\delta_1 >0$, $\exists$ an integer $N$ such that $m \ge N$ implies
\begin{eqnarray} \label{conv}
 r \frac{r^{-\alpha}\gamma}{1+r^{-\alpha}\gamma}\left|  \Lambda_m^P(r, \theta)- \Lambda(r, \theta)  \right| \le \quad \nonumber \\ 
 \max \left[ r \frac{r^{-\alpha}\gamma}{1+r^{-\alpha}\gamma}  \right] \left|  \Lambda_m^P(r, \theta)- \Lambda(r, \theta)  \right|  < \delta_1
\end{eqnarray}
which shows that $ \frac{r^{-\alpha}\gamma}{1+r^{-\alpha}\gamma}  \Lambda_n^P(r, \theta) $ is uniformly convergent to $\frac{r^{-\alpha}\gamma}{1+r^{-\alpha}\gamma}  \Lambda(r,\theta)$ on $[0,R]$, and $[0,\infty]$ as $\Lambda_m^P(r,\theta)$ and $\Lambda(r,\theta)$ are equal in $(R,\infty)$. Consequently, we can move the limit outside the integrals in the following expression resulting in
\begin{eqnarray}
\psi(\gamma) &=& \int_0^\infty\int_0^{2\pi} \lim_{m \to \infty}  \Lambda_m^P(r, \theta)r\frac{r^{-\alpha}\gamma}{1+r^{-\alpha}\gamma}d\theta dr \nonumber \\
 &=& \lim_{m \to \infty}\psi_m^P(\gamma).
\end{eqnarray}
Moreover, we express \eqref{FNP} as $F_m^P(\gamma) = g(\psi_m^P(\gamma))$, and $F(\gamma) = g(\psi(\gamma))$. Since $g(\psi_m^P(\gamma))$ is a continuous function of $\psi_m^P(\gamma)$, as $m \to \infty$, $\psi_m^P(\gamma) \to \psi(\gamma)$, which implies
\begin{eqnarray}
&&\lim_{m \to \infty}F_m^P(\gamma) = F(\gamma)\nonumber \\ 
&&= 1-\sum_{i=0}^{L-1}\frac{(\psi(\gamma)+\sigma^2\gamma)^i}{i!}\exp(-\psi(\gamma)-\sigma^2\gamma) \quad \quad \quad
\end{eqnarray}
which completes the proof.

\bibliography{main}

\end{document}